\documentclass[12pt]{article}
\usepackage{epsf}
\usepackage{graphicx}
\usepackage{amsfonts}
\usepackage[mathscr]{eucal}
\usepackage{booktabs}

\def\href#1#2{#2}   
\newif\ifdraft
\draftfalse	  
\reversemarginpar   
\makeatletter
\let\mlabel=\label
\let\adkendequation=\endequation%
\def\endequation{\adkendequation\adklabel\global\@ignoretrue}
\let\adkendeqnarray=\endeqnarray%
\def\endeqnarray{\adkendeqnarray\adklabel\global\@ignoretrue}
\newbox\marglabbox
\def\adklabel{\ifvoid\marglabbox\else\marginpar{\unhbox\marglabbox}\fi}
\def\label#1{\ifdraft\ifmmode%
  \global\setbox\marglabbox=\hbox{\hfill\fbox{\tiny\verb*~#1~}}%
  \else\ifinner\else\marginpar{\hfill\fbox{\tiny\verb*~#1~}}%
  \fi\fi\fi \mlabel{#1}}
\makeatother
\ifdraft%
\fi
\def\eusm{\mathscr}
\font\twelvefrak=eufm10 scaled 1200
\font\tenfrak=eufm10
  \newfam\frakfam
  
  \textfont\frakfam=\twelvefrak
  \scriptfont\frakfam=\tenfrak
  \scriptscriptfont\frakfam=\scriptfont\frakfam
\def\sqr#1#2{{\vcenter{\hrule height.#2pt
   \hbox{\vrule width.#2pt height#1pt \kern#1pt
      \vrule width.#2pt}
   \hrule height.#2pt}}}

\def\bsqr#1#2{{\vrule width #1pt height#2pt}}
\def\bsquare{{\mathchoice\bsqr66\bsqr66\bsqr33\bsqr33}}
\def\badbreak{\penalty1000}

\def\Trs{\mathop{\rm tr}}		    
\def\rational#1#2{{\mathchoice{\textstyle{#1\over#2}}%
  {\scriptstyle{#1\over#2}}{\scriptscriptstyle{#1\over#2}}{#1/#2}}}
\def\half{\rational12}			    
\newcommand{\cO}{{\cal O}}                  
\newcommand{\cP}{{\cal P}}                  
\newcommand{\tF}{{\tilde F}}                
\def\rpc{x}                                 
\def\Xg{X}                                  
\def\Fg{{\eusm X}}                          
\def\xd{P}                                  
\def\plp{{\Gamma}}                          
\def\cop{{C}}                               
\def\df{{\cP}_f}                            
\def\db{{\cP}_b}                            
\def\dop{{\xd}_r}                           
\def\Fgr{{\Fg}_r}                           
\begin{document}

\begin{center}
{\Large{\bf How Self--Dual is QCD?}} \\
\vspace*{.24in}
{\large{Andrei Alexandru$^1$ and Ivan Horv\'ath$^2$}}\\
\vspace*{.24in}
$^1$The George Washington University, Washington, DC, USA\\
$^2$University of Kentucky, Lexington, KY, USA

\vspace*{0.15in}
{\large{Oct 12 2011}}

\end{center}

\vspace*{0.05in}

\begin{abstract}

  \noindent
  Vacuum characteristics quantifying {\em dynamical} tendency toward self--duality 
  in gauge theories could be used to judge the relevance of classical solutions or 
  the viability of classically motivated vacuum models. Here we decompose the field 
  strength of equilibrium gauge configurations into self--dual and anti--self--dual parts, 
  and apply absolute $\Xg$--distribution method to the resulting polarization dynamics 
  in order to construct such characteristics. Using lattice regularization and focusing on 
  pure--glue SU(3) gauge theory at zero temperature, we find evidence for positive
  but very small dynamical tendency for self--duality of vacuum in the continuum limit. 

\end{abstract}

\vspace*{0.10in}

\noindent{\bf 1. Introduction.}
The availability of topologically non--trivial classical solutions to Yang--Mills 
equations~\cite{ins_disc}, combined with technology of semiclassical 
calculations, provided the framework for discussion of important non--perturbative 
effects such as the $U_A(1)$ problem of QCD~\cite{ins_tHooft}, 
$\theta$--dependence~\cite{ins_theta} or the baryon number violation in the Standard 
Model~\cite{ins_tHooft}. However, the merits of semiclassical approximations have 
been questioned since the early days, especially at zero temperature and away from 
the Higgs phase~\cite{WittenUA(1)}. 

The applicability of semiclassical approach is related to vacuum properties of the  
theory in question. Indeed, if classical solutions represent a sensible starting point of 
a systematic analysis, then it is expected that their properties will manifest themselves 
in configurations dominating the associated path integral. In that vein, the semiclassically 
motivated picture of QCD vacuum based on instantons has been evolving~\cite{ins_gas,ILM}, 
focusing mainly on the physics of spontaneous chiral symmetry breaking~\cite{ins_mixing} 
and with confinement being rather problematic. In fact, the large N--motivated arguments 
of Witten~\cite{WittenUA(1)} qualitatively amount to the proposition that, in the confining 
vacuum, the condition of field being pure gauge on a boundary enclosing 4--dimensional 
subvolume cannot be satisfied even approximately. This implies that the topological charge 
of true QCD vacuum is not expected to be locally quantized in near--integer values as 
predicted by instanton--based picture. The absence of this local quantization is 
consistent with direct lattice QCD calculations~\cite{Hor02B}. Moreover, it has been 
found~\cite{Hor03A} and confirmed~\cite{Ilg07A} that the topological charge in pure--glue 
QCD vacuum organizes into a lower--dimensional sign--coherent global structure. This signals 
vacuum arrangement of very different nature than the one suggested by semiclassical reasoning. 

It would be convenient and desirable to have a vacuum--based measure at our disposal, capable 
of probing the ``degree of classicality'' in the theory. What we have in mind is a simple 
characteristic, computable by lattice methods, that could signal the basic expectation 
for validity of semiclassical approximations and the semiclassical picture of the vacuum. 
Possible strategy for such an approach has been put forward in Ref.~\cite{Hor01A}, but it has 
not been properly carried out. In particular, for gauge theories, classicality is intimately 
connected to (anti)self--duality. Indeed, self--duality of a field configuration suffices 
to solve the equations of motion in pure--glue SU(N) gauge theory 
(see e.g.~\cite{Cre10A} for recent review), and the true local minima 
of the action are of this type. Thus, in practical terms, self--duality of vacuum 
configurations is almost synonymous with what we wish to characterize.

In their initial study of self--duality, the authors of Ref.~\cite{Hor01A} pursued 
an {\em indirect} approach via Dirac eigenmodes. In particular, they proposed a differential 
characteristic of local 
chiral behavior in the near--zero modes ($\Xg$--distribution), which could also be viewed as 
a low--energy reflection of possible self--duality in the gauge field. In Ref.~\cite{Gat02A} 
the same technique has then been used in conjunction with a particular form of truncated 
eigenmode expression for field strength itself. However, the crucial shortcoming of the above 
attempts is that they are arbitrary with respect to the choice of the polarization measure. 
Indeed, as demonstrated in Ref.~\cite{Ale10A}, it is possible to obtain any desired behavior of 
$\Xg$--distribution by the appropriate adjustment of the reference frame (polarization function). 
Thus, the method in this form is inherently kinematical.\footnote{In fact, as first shown 
in Ref.~\cite{Dra04A}, the chiral polarization effects observed in works that followed 
Ref.~\cite{Hor01A} and aimed at studying the merits of instanton liquid model of QCD 
vacuum~\cite{Followup}, are almost entirely due to kinematics.}

This issue has been recently addressed in detail, and resulted in the general polarization method 
of {\em absolute $\Xg$--distribution}~\cite{Ale10A}. The polarization measure is uniquely fixed 
in this framework, so as to provide a differential comparison of polarization tendencies in question 
relative to those involving statistically independent polarization components. By virtue of being 
a valid polarization measure and, at the same time, a comparator to statistical independence, 
absolute $\Xg$--distribution $P_A(\Xg)$ represents a {\em dynamical} polarization characteristic. 
The associated integrated quantity, the correlation coefficient of polarization 
$\cop_A$ ($-1\le \cop_A \le 1$), expresses the overall dynamical tendency for polarization. 

One noteworthy virtue of adopting correlational characteristics is that it alleviates the usual 
worries about distortions of vacuum structure due to ``ultraviolet noise''. These worries typically 
result in attempts to suppress high frequency fluctuations, and were in fact part of the reason 
for focusing on low--lying Dirac modes in the above context of self--duality. However, with
absolute polarization methods, this is not necessary and it is appropriate to approach the problem 
directly 
in equilibrium configurations of the gauge field thus capturing the properties of full dynamics.
Indeed, if ultraviolet fluctuations represent pure noise, they will not affect the direction
of dynamical tendencies. On the other hand, if these fluctuations exhibit correlations, then 
there is no reason to suppress them since they can affect the physics, even at low energy. 

Given the above, we propose the correlation coefficient of polarization $\cop_A$ 
(and the underlying absolute $\Xg$--distribution), evaluated for the duality decomposition 
of the gauge field strength tensor, as a suitable vacuum measure of self--duality, and thus a candidate 
for computationally accessible characteristic of classicality in 4--dimensional gauge theories. 
Utilizing the method, we show that pure--glue QCD exhibits very small positive dynamical tendency 
for self--duality. Specifically, using Wilson lattice definition of the theory and overlap--based 
definition of the field strength tensor, we find that the absolute $\Xg$--distribution is mildly 
convex with $\cop_A \approx 0.018$. 
This means that in head-to-head comparisons of samples chosen from QCD--generated field strength and 
from ensemble of its statistically independent duality components, the probability for the former 
being more self--dual is only by $\cop_A/2 \approx 0.009$ larger than equal chance. We also 
construct the effective low energy field strength based on overlap eigenmode expansion, which is found 
to have comparable degree of dynamical self--duality.

\bigskip

\noindent{\bf 2. The Absolute $\Xg$--Distribution Method.}
We begin with a brief description of the absolute $\Xg$--distribution method. Detailed discussion 
is given in Ref.~\cite{Ale10A}. Consider a quantity $Q$ with values in some vector space that can be 
decomposed into a fixed pair of equivalent orthogonal subspaces, i.e. $Q = Q_1 + Q_2$ with 
$Q_1 \cdot Q_2 = 0$. The dynamics of $Q$ is governed by the probability distribution $\df(Q_1,Q_2)$ 
which is assumed to be symmetric with respect to the two subspaces, i.e. $\df(Q_1,Q_2)=\df(Q_2,Q_1)$. 
In field theory applications the distribution $\df(Q_1,Q_2)$ descends from the distribution of field 
configurations defined by the local action.

We are interested in characterizing the property of polarization, namely the tendency for asymmetry 
in the contribution of the two subspaces in the preferred values of $Q$. The contribution of each 
subspace in sample $Q$ is measured by magnitudes $q_i \equiv |Q_i|$ of their components, and one 
can thus reduce the full distribution $\df(Q_1,Q_2)$ to the distribution of magnitudes $\db(q_1,q_2)$ 
for this purpose. Moreover, the asymmetry in the contribution of the two subspaces in $Q$ can be 
assessed by the ratio $q_2/q_1$ which can be cast into a normalized variable by considering
a combination such as~\cite{Hor01A}
\begin{equation}
    \rpc \,=\, \frac{4}{\pi} \, \tan^{-1}\Bigl(\frac{q_2}{q_1}\Bigr) \,-\,1   
          \equiv \Fgr(q_1,q_2) \qquad \quad
    \rpc \in [-1,1]
    \label{eq:100}
\end{equation}
namely the {\em reference polarization coordinate}. Notice that $\rpc=-1$ if sample $Q$ is 
strictly polarized in first direction, $\rpc=1$ if it is strictly polarized in second direction, and
$\rpc=0$ for strictly unpolarized sample $q_1=q_2$.  The marginal distribution of reference 
polarization coordinate in distribution $\db(q_1,q_2)$, namely
\begin{equation}
    \dop(\rpc) \,=\,  
      \int_0^\infty d q_1 \int_0^\infty d q_2 \,
      \db(q_1,q_2) \; \delta\Bigl( \rpc - \Fg_r(q_1,q_2) \Bigr)
      \label{eq:110}
\end{equation}
is the {\em reference \Xg--distribution}~\cite{Hor01A}. It can be viewed as detailed but kinematical
polarization characteristic of $\df(Q_1,Q_2)$ since its qualitative properties depend on 
the choice of $\Fgr$.

Rather than the distribution of reference polarization function (coordinate) $\Fg_r(x)=x$,
the {\em absolute $\Xg$--distribution} $\xd_A(\Xg)$ represents a distribution of 
the polarization function
\begin{equation}
   \Fg_A(\rpc) \equiv 2 \int_{-1}^\rpc d y \, \dop^u(y) - 1 
   \label{eq:120}   
\end{equation}
where the reference $\Xg$--distribution $\dop^u(\rpc)$ is computed for the dynamics of 
statistically independent components, namely 
\begin{equation}
    \db^u(q_1,q_2) \,\equiv\, p(q_1)\, p(q_2)   \qquad\quad
    p(q) \equiv  \int_0^\infty d q_2 \, \db(q,q_2) \,=\,
                      \int_0^\infty d q_1 \, \db(q_1,q)
   \label{eq:130}
\end{equation}
There are three important points to note about $\xd_A(\Xg)$ so constructed. 
{\em (i)} $\Fg_A(\rpc)$ is a valid polarization measure in exactly the same sense as $\rpc$ 
itself is: it quantifies the degree of polarization in the sample. 
{\em (ii)} As one can easily check, the distribution of $\Fg_A(\rpc)$ is uniform 
in uncorrelated distribution $\db^u(q_1,q_2)$ and thus the absolute $\Xg$--distribution directly 
measures polarization tendencies relative to statistical independence. One can best see 
the differential nature of this direct comparison from the explicit form 
\begin{equation}
   \xd_A(\Xg) \,=\, \frac{1}{2} \, 
                  \frac{\dop   \Bigl(  \Fg_A^{-1}\,(\,\Xg\,) \Bigr)}
                       {\dop^u \Bigl(  \Fg_A^{-1}\,(\,\Xg\,) \Bigr)}
   \label{eq:140}
\end{equation}
{\em (iii)} The distribution $\xd_A(\Xg)$ is invariant under the choice of reference polarization
function (coordinate) used in its construction, and is in this sense absolute. 

Taken together, properties {\em (i--iii)} indicate that the absolute $\Xg$--distribution represents
a {\em dynamical} polarization characteristics. This differential information can be reduced to
integrated form by computing the associated correlation coefficient of polarization, namely
\begin{equation}
   \cop_A \,\equiv \,
    2\,\int_{-1}^{1} d \Xg \, |\Xg| \, \xd_A(\Xg) \,-\,1
    \label{eq:150}
\end{equation}
The positive value of $\cop_A \in [-1,1]$ indicates that the dynamics enhances polarization 
relative to statistical independence, while the negative value implies its dynamical suppression.
Indeed, the precise statistical meaning of $\cop_A$ is as follows. Consider the experiment
wherein samples are independently drawn from $\db(q_1,q_2)$ and 
$\db^u(q_1,q_2)$ in order to be compared, head to head, with respect to their degree 
of polarization. The result of this experiment is the probability $\plp_A$ that a sample 
drawn from ensemble in question is more polarized than sample drawn from ensemble of 
statistically independent components. Then $\plp_A=(\cop_A + 1)/2$ thus implying the above
interpretation of $\cop_A$.

\bigskip

\noindent {\bf 3. Duality Decomposition.}
Our object of interest is the Euclidean field-strength tensor $F(x)$ 
of SU(N) gauge theory in four dimensions. To analyze its dynamical polarization 
properties with respect to duality, we decompose it into its self--dual and 
anti--self--dual parts, i.e.
\begin{equation}
   F \,=\, F_S + F_A    \qquad\; \mbox{\rm with} \qquad\;  
   F_S \equiv \half (F + \tF)   \qquad F_A \equiv \half (F - \tF) 
\end{equation}
where the dual tensor $\tF$ is defined in a standard way, namely
\begin{equation}
   \tF_{\mu\nu} \,\equiv\, \half \epsilon_{\mu\nu\rho\sigma} F_{\rho\sigma}
   \qquad\quad  \epsilon_{1234} \equiv 1
\end{equation}
The above split represents an orthogonal decomposition of the field--strength
tensor relative to the scalar product
\begin{equation}
   A \cdot B \equiv \half \sum_{\mu\nu} \Trs A_{\mu\nu} B_{\mu\nu} 
\end{equation}
with $A\equiv\{A_{\mu\nu}\}$, $B\equiv\{B_{\mu\nu}\}$ representing arbitrary
antisymmetric collections of $N \times N$ Hermitian matrices. 
Thus, to evaluate dynamical tendency for self--duality in SU(N) gauge theory
we apply absolute polarization methods to objects $(Q_1,Q_2) \equiv (F_S,F_A)$, 
which in practice involves statistical analysis of pairs 
$(q_1,q_2) \leftrightarrow (|F_S|,|F_A|)$ with the norm defined by 
the scalar product above.

\bigskip

\noindent{\bf 4. Lattice QCD Setup.}
Our exploratory calculations will be performed in the context of pure glue SU(3) lattice
gauge theory with Iwasaki action~\cite{Iwasaki}. Table~\ref{tab:ensembles} summarizes 
the parameters of the ensembles. Lattice scale has been determined from the string tension 
following the methods and results of Ref.~\cite{Oka99A}. Ensembles $E_2$, $E_3$, $E_8$, 
$E_4$ and $E_7$ have the same physical volume, and serve to investigate the continuum limit. 
Ensemble $E_6$ has larger physical volume to check for possible finite size effects.

\newcommand\fm{\mathop{\rm fm}}
\begin{table}[b]
   \centering
   \begin{tabular}{@{} ccccccc @{}} 
      \toprule
      Ensemble  &  Size  &  Volume  &  Lattice Spacing  &  Iwasaki $\beta$ & 
      $N_{\rm config}^F$  &  $N_{\rm config}^{\rm eigen}$\\
      \midrule
     $E_{2}$  &  $12^{4}$  & $(1.32\fm)^{4}$  &  $0.110  \fm$ &  $2.530$  & 400  & 97\\
     $E_{3}$  &  $16^{4}$  & $(1.32\fm)^{4}$  &  $0.0825 \fm$ &  $2.725$  & 200  & 99\\
     $E_{8}$  &  $20^{4}$  & $(1.32\fm)^{4}$  &  $0.066  \fm$ &  $2.892$  &  80  &   \\
     $E_{4}$  &  $24^{4}$  & $(1.32\fm)^{4}$  &  $0.055  \fm$ &  $3.0375$ &  39  & 96\\
     $E_{7}$  &  $32^{4}$  & $(1.32\fm)^{4}$  &  $0.041  \fm$ &  $3.278$  &  19  &   \\
     $E_{6}$  &  $32^{4}$  & $(1.76\fm)^{4}$  &  $0.055  \fm$ &  $3.0375$ &  20  & 20\\
      \bottomrule
   \end{tabular}
   \caption{Ensembles used in the overlap calculations of $F_{\mu\nu}(x)$. Column 
   $N_{\rm config}^F$ shows the number of configurations for which local field 
   strength (\ref{eq:410}) has been computed. Column $N_{\rm config}^{\rm eigen}$
   lists the number of configurations with computed eigenmodes when available.}
   \label{tab:ensembles}
\end{table}

We will utilize the local lattice definition of gauge field strength tensor based 
on the relation~\cite{Hor06B,Liu07A,Ale08A}
\begin{equation}
   {\Trs}_s\, \sigma_{\mu\nu} \, D_{0,0}(U(a)) 
   \;=\; c^T\, a^2\, F_{\mu\nu}(0) \;+\; \cO(a^4)
   \label{eq:410}
\end{equation} 
with $c^T\equiv c^T(\rho,r)$ being generically non--zero for overlap Dirac 
operator~$D\equiv D^{\rho,r}$~\cite{Neu98BA}. The above expression is
valid for discretization $U(a)$ of arbitrary classical field $A_\mu$, 
and ${\Trs}_s$ denotes trace over spin indices only. Parameters $r=1$ and $\rho=26/19$ 
were used for calculations throughout this paper. Local values of $F_{\mu\nu}(x)$ in 
equilibrium lattice QCD configurations were obtained by evaluating the needed 
matrix elements of $D$ using point sources.

One advantage of the above definition for the field strength is that it can be naturally
eigenmode--expanded to provide an effective low--energy representation. To do that,
we follow the same strategy as one used in definition of the effective topological
density in Ref.~\cite{Hor02B}. In particular, rather than truncating the expression 
for lattice field implied by (\ref{eq:410}), we take advantage of the fact that 
$\Trs \sigma_{\mu\nu} = 0$ and rather expand
\begin{equation}
   F_{\mu\nu}(x) \equiv \frac{1}{c^T} \, {\Trs}_s \, \sigma_{\mu\nu} \, D_{x,x}   
   \,=\, -\frac{1}{c_T} \, {\Trs}_s \, \sigma_{\mu\nu} \, ( 2\rho - D_{x,x} )   
   \label{eq:420}
\end{equation}
Indeed, while the expansion of the former gives lowest weight to low--lying modes
and highest weight to the modes at the cutoff, the latter reverses this and gives
highest modes zero weight as naturally desired. The effective field at the fermionic 
scale $\Lambda$ is then
\begin{equation}
   \Bigl [ F_{\mu\nu}^{\Lambda}(x) \Bigr]_{a,b} \equiv 
   -\frac{1}{c_T} \, \sum_{|\lambda| \le \Lambda a} ( 2\rho - \lambda) 
   \Bigl[ \psi_\lambda(x) \Bigr]_b^\dagger \sigma_{\mu\nu} \Bigl[ \psi_\lambda(x) \Bigr]_a
   \label{eq:430}
\end{equation}
where $\psi_\lambda(x)$ is the eigenmode of $D$ with eigenvalue $\lambda$ and $a,b$ are 
color indices.

Low--lying eigenmodes needed for the construction of effective field strength 
were computed using our own implementation of the implicitly restarted Arnoldi 
algorithm with deflation~\cite{ira}. The zero modes and about 50 pairs of lowest near--zero 
modes with complex--conjugated eigenvalues were obtained for each configuration. 
In Table~\ref{tab:ensembles} we indicate the ensembles with eigenmodes available
for these purposes. One point to note here is that, as one can easily check, 
the terms associated with chiral zero modes are strictly self--dual. While the relative 
weight of this contribution is zero both in the infinite volume limit at fixed lattice 
spacing, and in the continuum limit at fixed physical volume, the zeromodes do contribute
in finite lattice calculations and are included.

\begin{figure}[t]
\begin{center}
    \centerline{
    \hskip 0.00in
    \includegraphics[width=16.0truecm,angle=0]{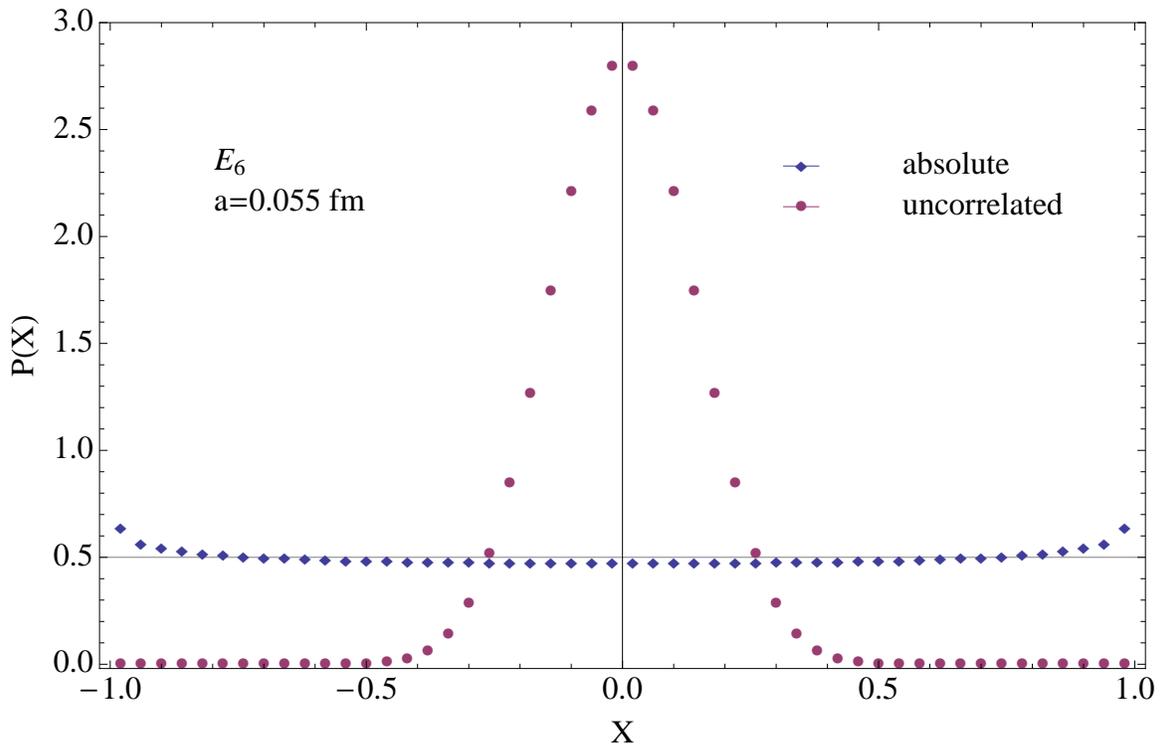}
     }
     \vskip -0.00in
     \caption{Absolute $\Xg$--distribution in duality and reference $\Xg$--distribution 
     of statistically independent duality components for ensemble $E_6$.}
     \label{fig:PAR_E6}
\end{center}
\end{figure} 

\bigskip

\noindent{\bf 5. Main Results.}
We have computed absolute $\Xg$-distributions in duality for all ensembles listed 
in Table~\ref{tab:ensembles}. The plot of Fig.~\ref{fig:PAR_E6} shows the result 
for ensemble $E_6$ which has the largest physical volume. Note that the associated kinematical 
background represented by the reference $\Xg$--distribution of statistically independent 
components is also shown.\footnote{As emphasized in Ref.~\cite{Ale10A}, the combination
of absolute $\Xg$--distribution and reference $\Xg$--distribution of uncorrelated
components provides all that is necessary to construct $\Xg$--distribution in arbitrary
reference frame (polarization function). In that sense, it stores a complete information
on polarization, dynamical and kinematical.}
As can be clearly seen, despite the completely ``unpolarized--looking'' kinematics, there is 
a slight dynamical tendency of the gauge field strength tensor to polarize itself. Indeed, 
the absolute $\Xg$--distribution shows a small excess of probability near the extremal values 
of absolute polarization coordinate. This observation represents 
the central message of this work since, as we discuss below, the data suggests that such 
dynamical behavior persists both in the continuum and infinite volume limits.

\begin{figure}                        
                                                                                
\begin{center}                                                                                                        
    \centerline{                                                                                                      
    \hskip 0.08in                                                                                                     
    \includegraphics[width=8.8truecm,angle=0]{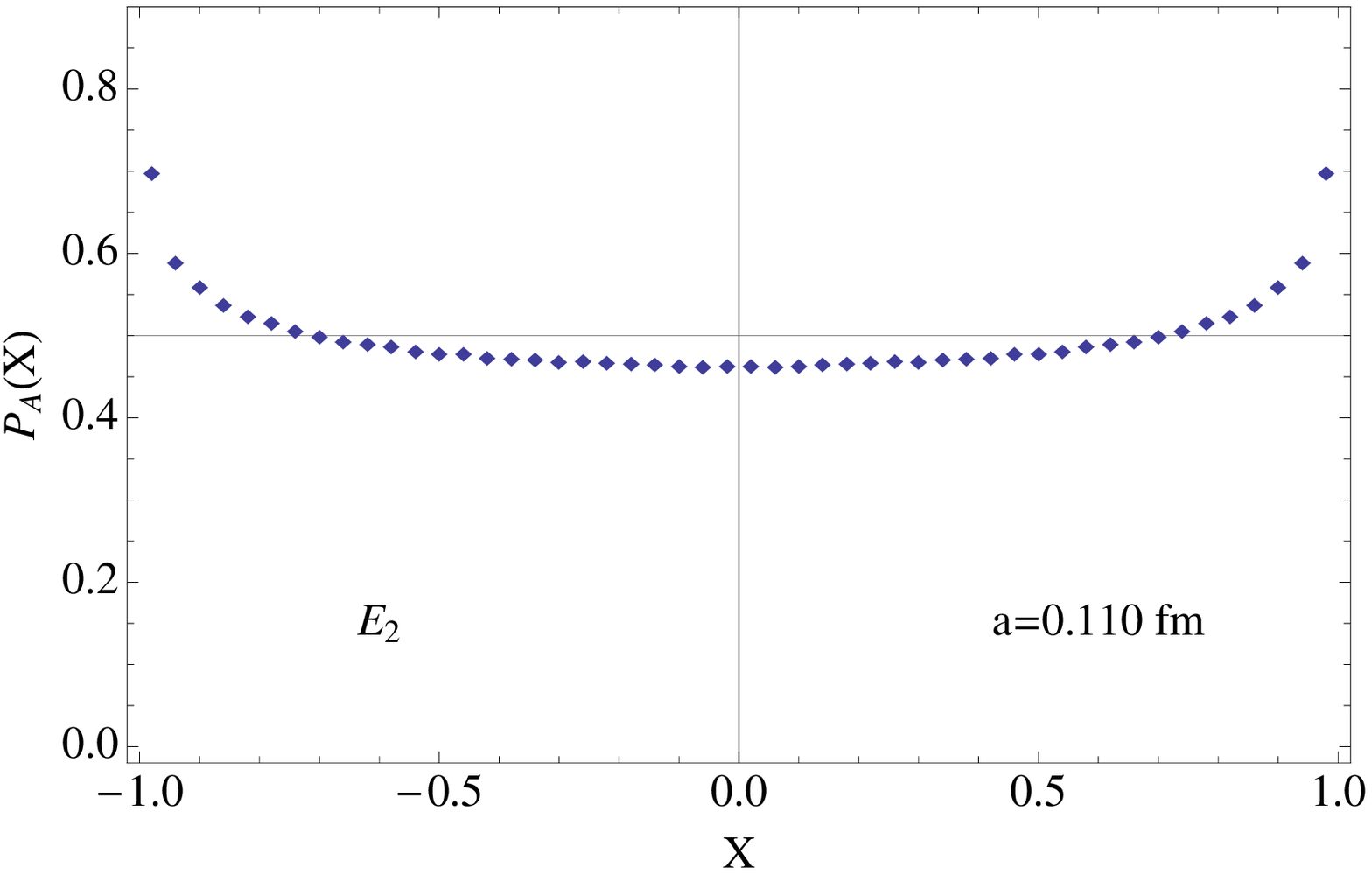}                                             
    \hskip -0.1in                                                                                                     
   \includegraphics[width=8.8truecm,angle=0]{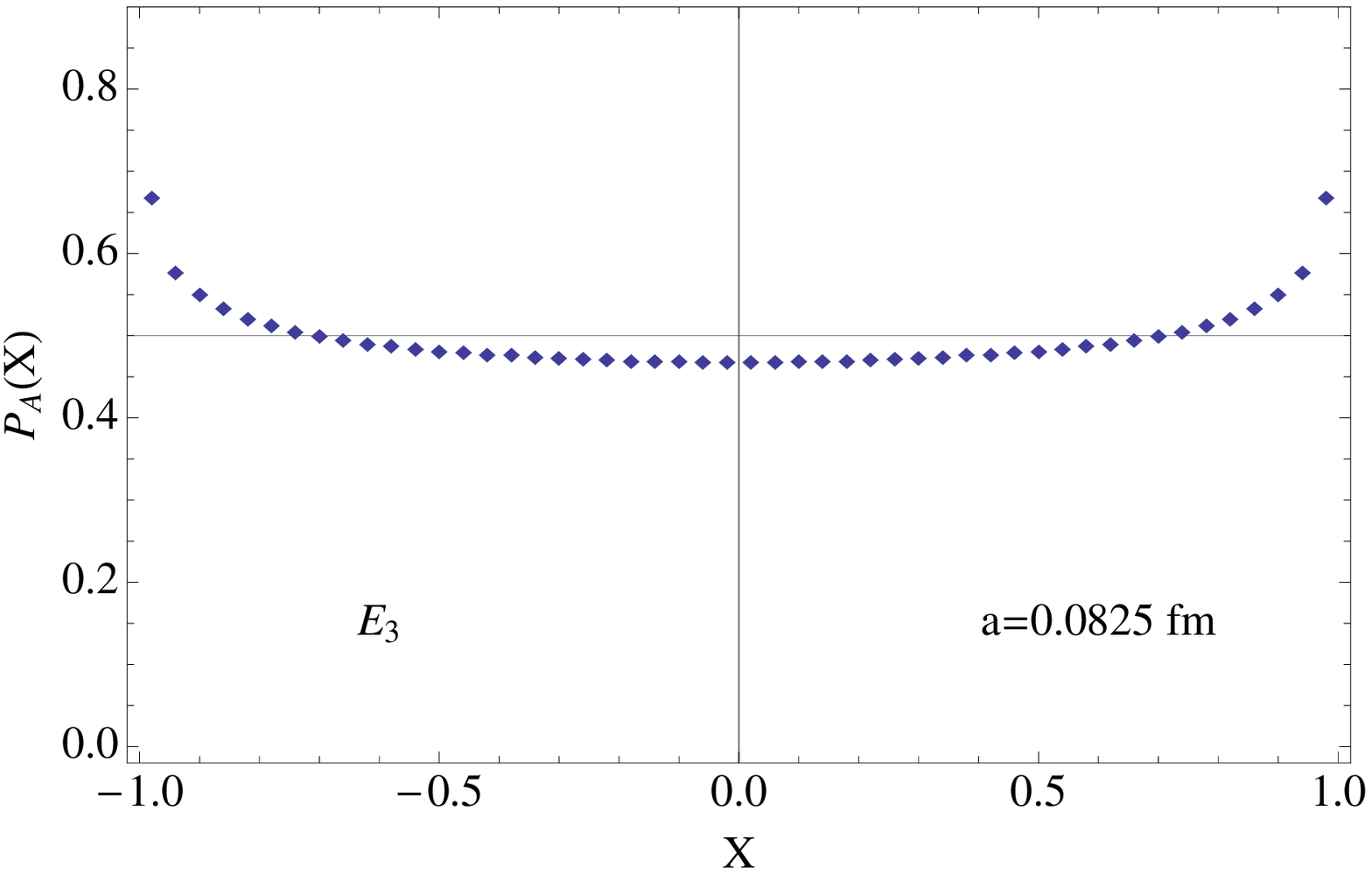}                                           
     }                                                                                                                
     \vskip 0.10in                                                                                                   
    \centerline{
    \hskip 0.08in
    \includegraphics[width=8.8truecm,angle=0]{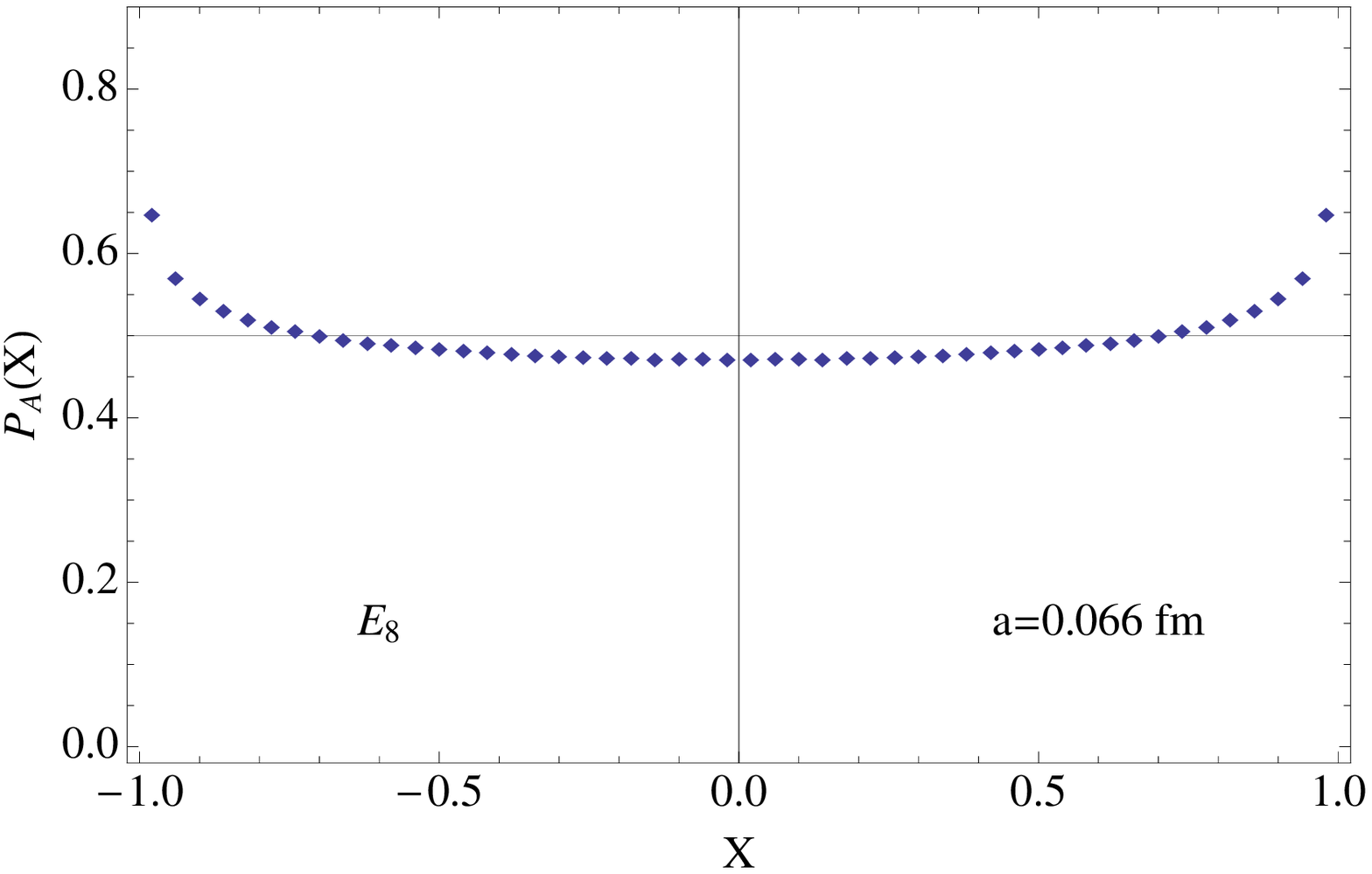}                                             
    \hskip -0.1in
    \includegraphics[width=8.8truecm,angle=0]{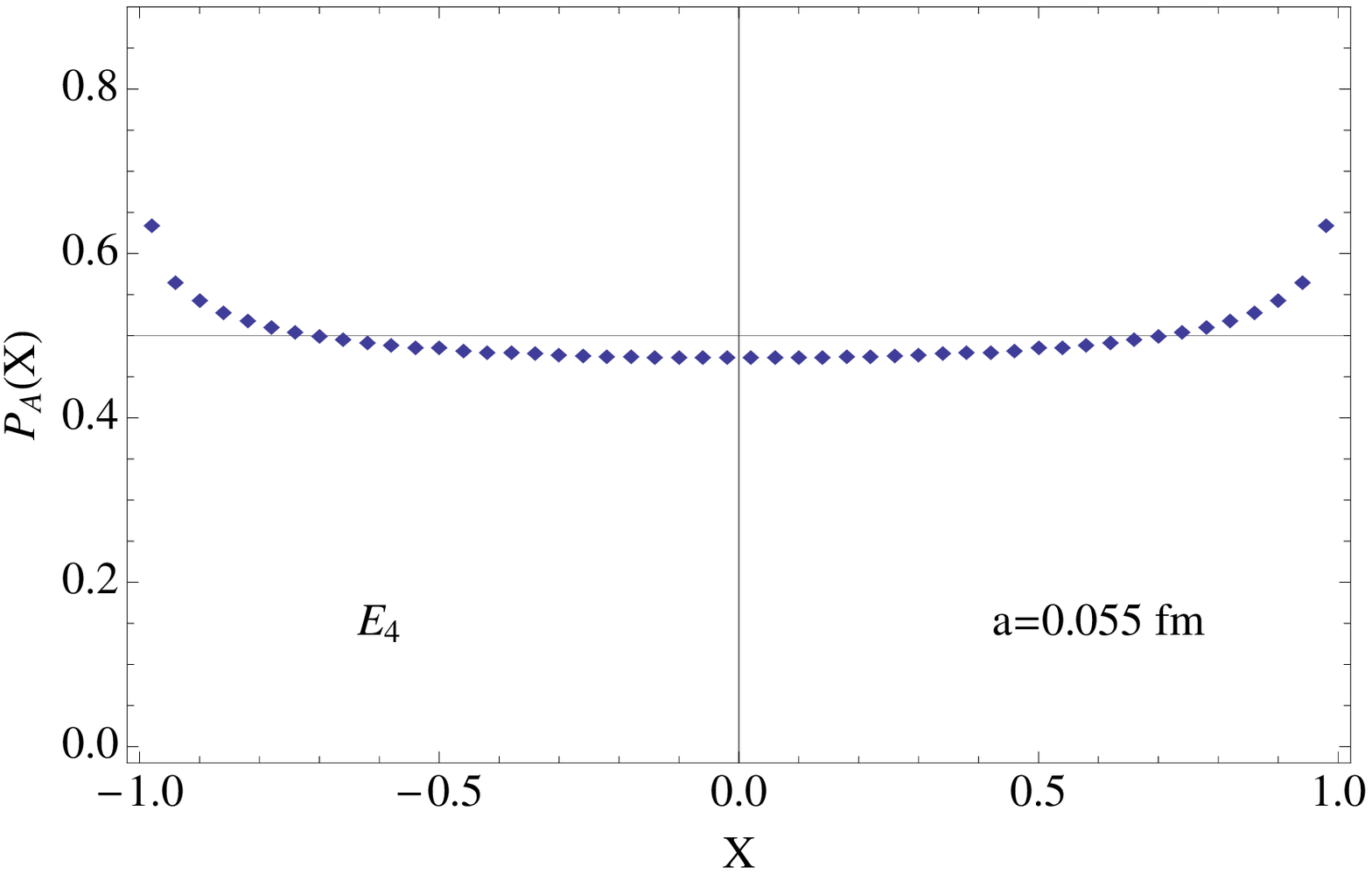}                                           
     }                                                                                                                
     \vskip 0.10in
    \centerline{
    \hskip 0.08in
    \includegraphics[width=8.8truecm,angle=0]{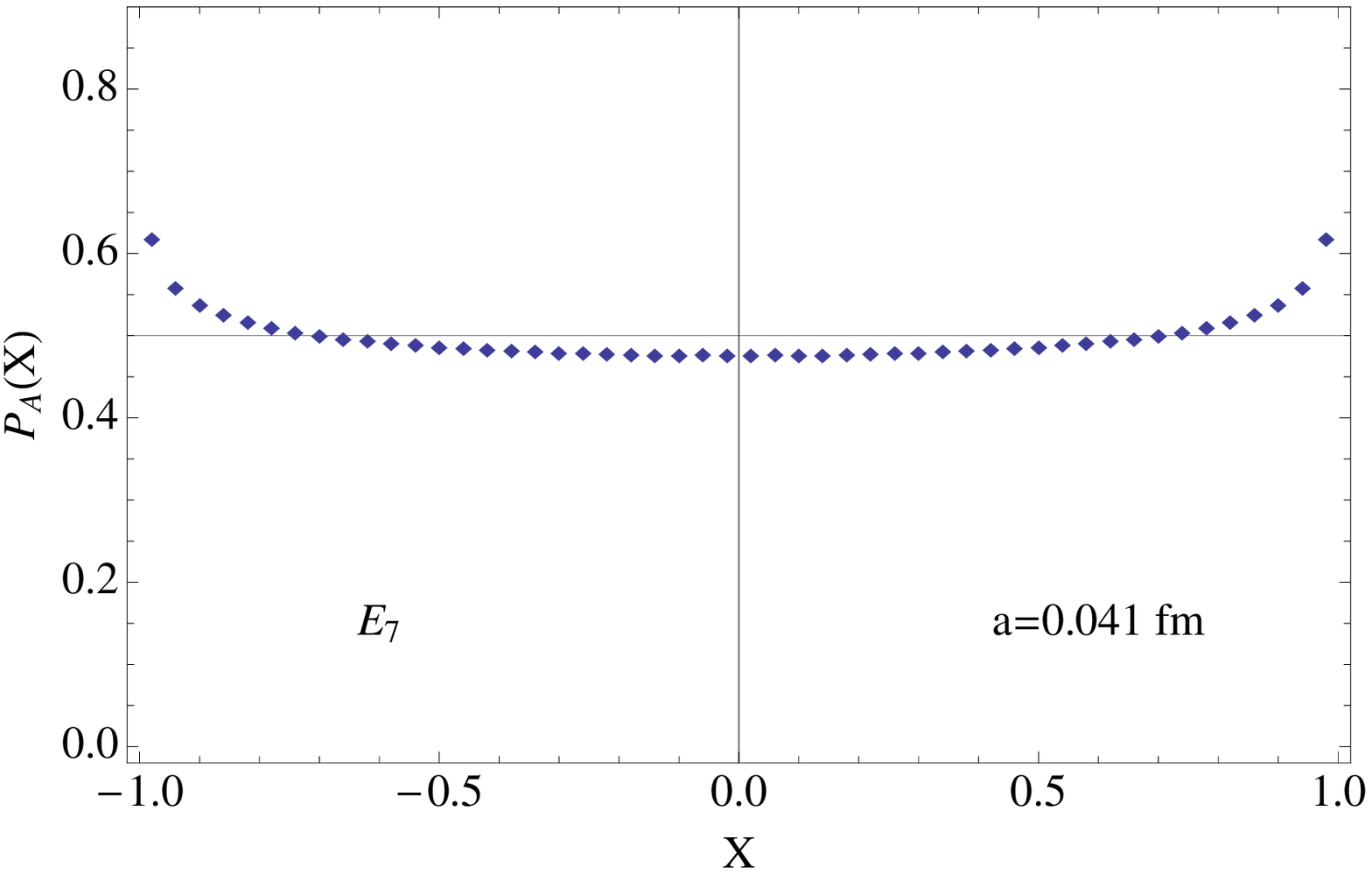}
    \hskip -0.1in
    \includegraphics[width=8.8truecm,angle=0]{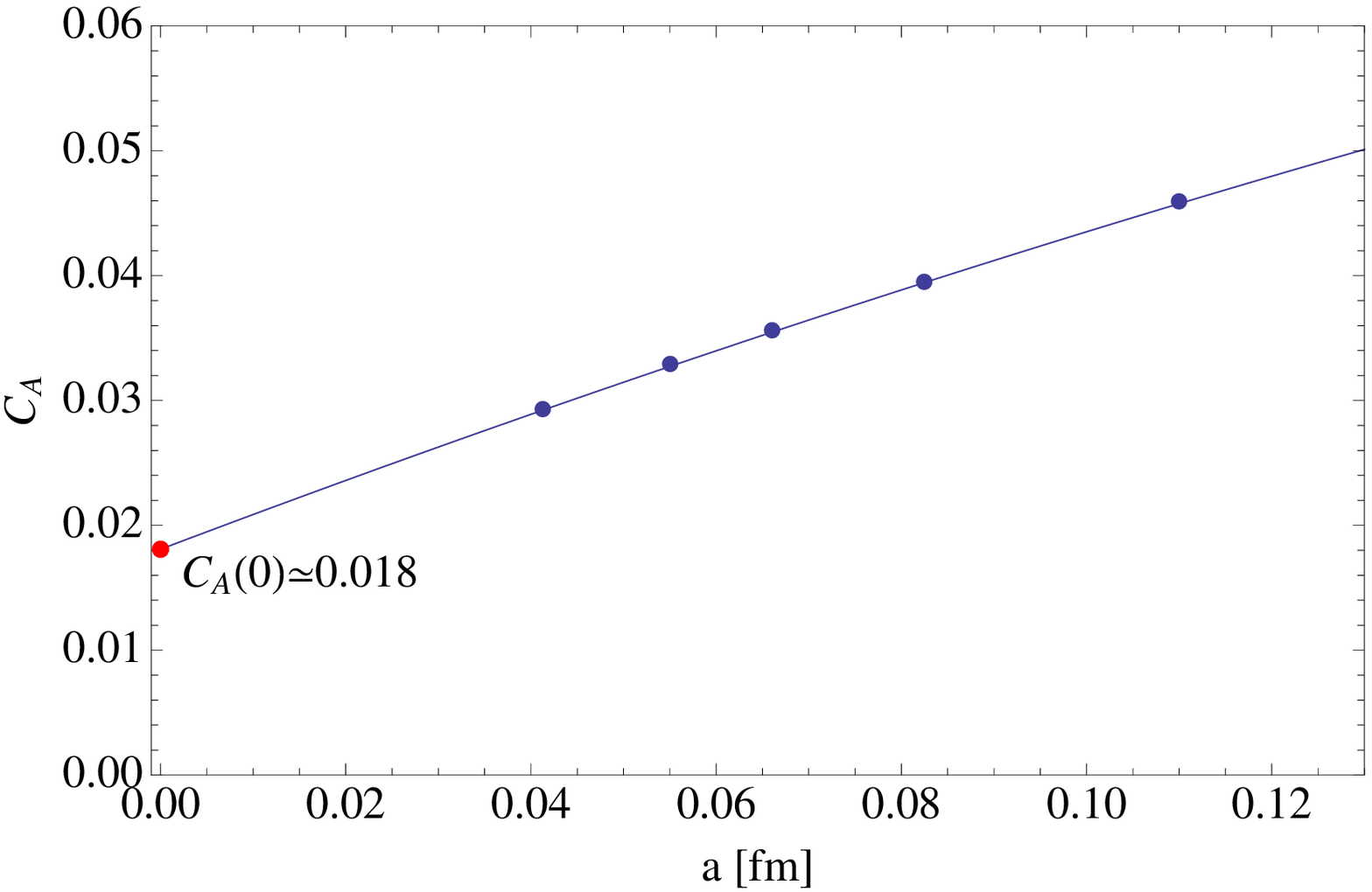}     
     }                                                                                                                
     \vskip 0.00in                                                                                                   
     \caption{Absolute duality $\Xg$--distributions for all ensembles at fixed
     physical volume $V=(1.32 \, {\mbox{\rm fm}})^4$. Lattice spacing is decreasing 
     in lexicographic order of plots. The last panel shows the corresponding lattice 
     spacing dependence of the polarization correlation coefficient, together 
     with the quadratic fit for extrapolation to the continuum limit. All data
     points in the above plots have errorbars too small to be clearly seen.}
    \label{fig:PA_all}
\end{center}
\end{figure}

Focusing on absolute $\Xg$-distributions from now on, we show them more closely for all
ensembles with fixed physical volume in Fig.~\ref{fig:PA_all}. 
As can be directly inspected, these distributions are all mildly convex, implying small 
positive dynamical tendency for polarization in duality. Despite the fact that this tendency 
slowly decreases toward the continuum limit, the data strongly suggests that a finite 
positive continuum limit does exist. In Fig.~\ref{fig:PA_all} we also show a fit to quadratic 
polynomial in lattice spacing, which we take to facilitate this continuum extrapolation.

\begin{figure}[t]
\begin{center}
    \centerline{
    \hskip 0.00in
    \includegraphics[width=15.0truecm,angle=0]{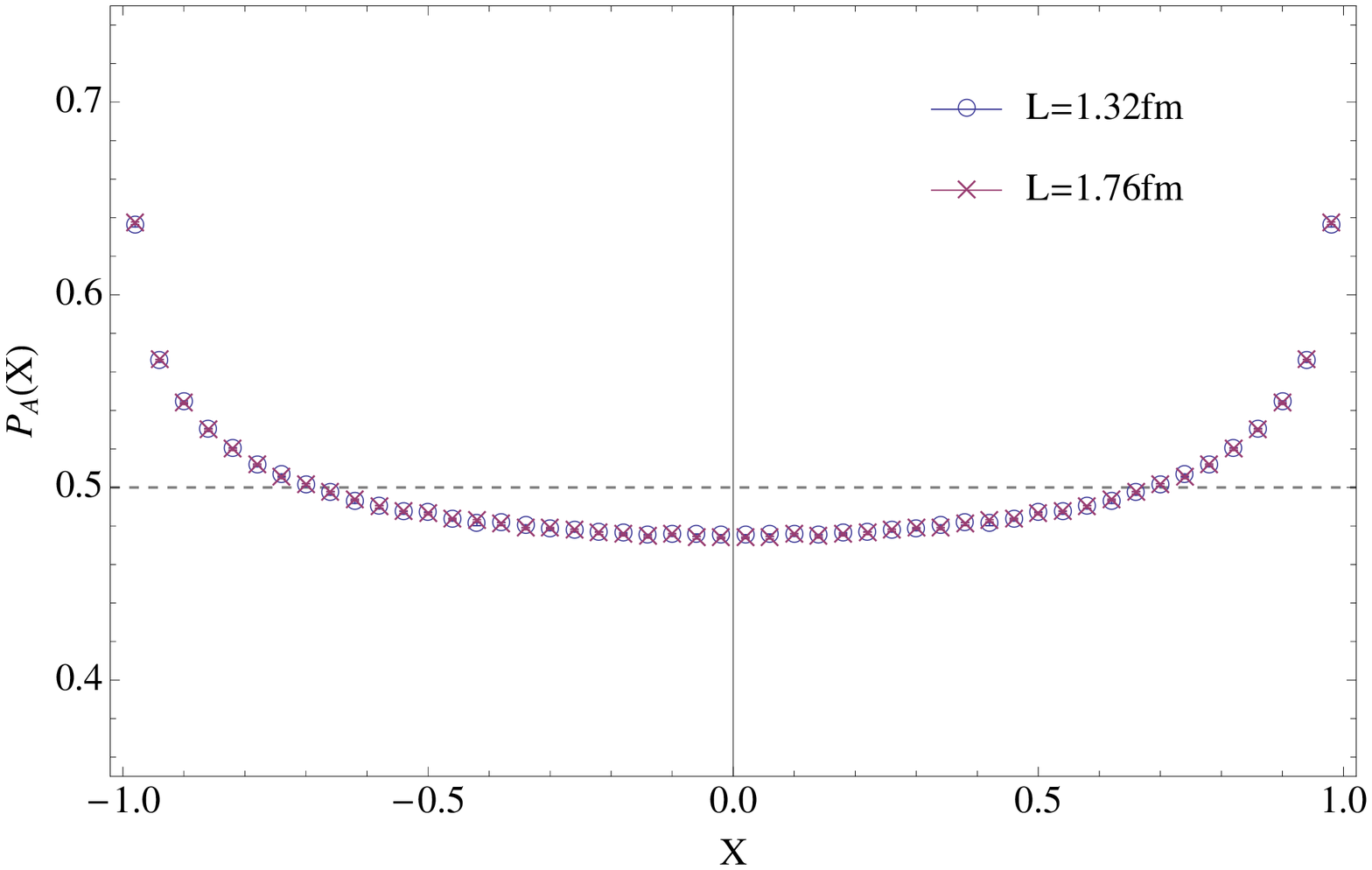}
     }
     \vskip -0.00in
     \caption{The comparison of absolute $\Xg$--distributions at the same lattice spacing
     but different volumes ($E_4$ and $E_6$).}
     \vskip -0.1in 
     \label{fig:PA_E4_E6}
\end{center}
\end{figure}

Although the physical volume in the above continuum extrapolation is rather small,
the finite volume corrections for observables at hand turn out to be negligible. 
Indeed, in Fig.~\ref{fig:PA_E4_E6} we show the comparison of absolute 
$\Xg$--distributions for ensembles $E_4$ and $E_6$ that have identical couplings 
($a=0.055$ fm) but with $E_6$ representing a significantly larger physical volume. 
As can be seen in this close--up view, the two distributions are rather difficult 
to distinguish one from another.

The above results make strong case for the following proposition that we suggest for further
investigation.

\medskip
\noindent {\em \underline{Proposition 1}:$\;$ Absolute $\Xg$--distributions for duality
           in SU(3) pure glue lattice gauge theory are convex. This property, and
           thus the associated positive tendency for (anti)self--duality, will persist
           in the continuum limit. }
\medskip

\noindent It needs to be emphasized that the observed polarization tendency, as objectively quantified 
by the associated correlation coefficient, is very small. In particular, $C_A \approx 0.018$ for 
continuum limit taken using Iwasaki lattice gauge action and overlap-based definition of 
the field--strength tensor.

\bigskip

\begin{figure}[t]
\begin{center}

    \centerline{
    \hskip 0.08in
    \includegraphics[width=8.8truecm,angle=0]{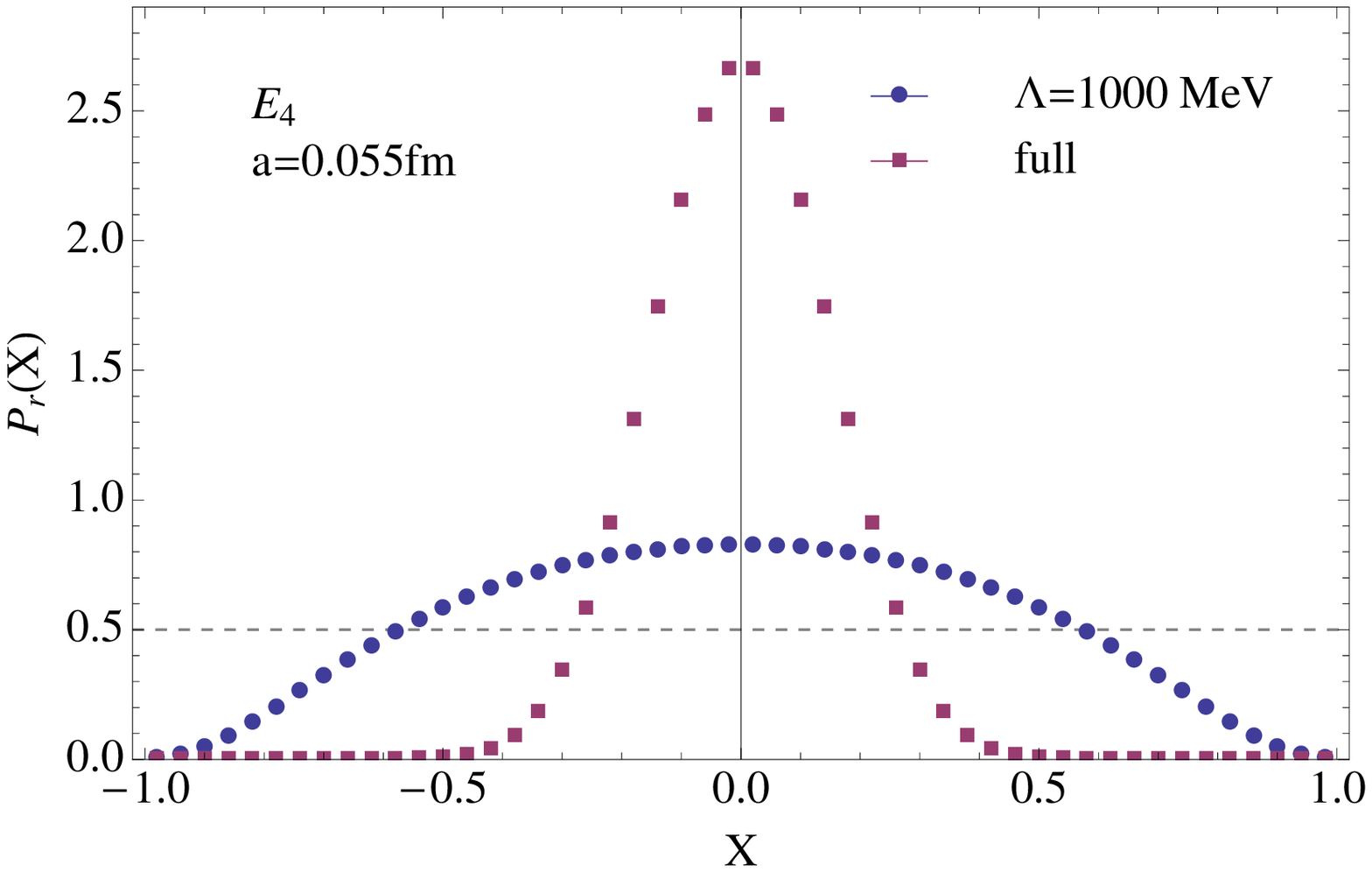}
    \hskip -0.1in
    \includegraphics[width=8.8truecm,angle=0]{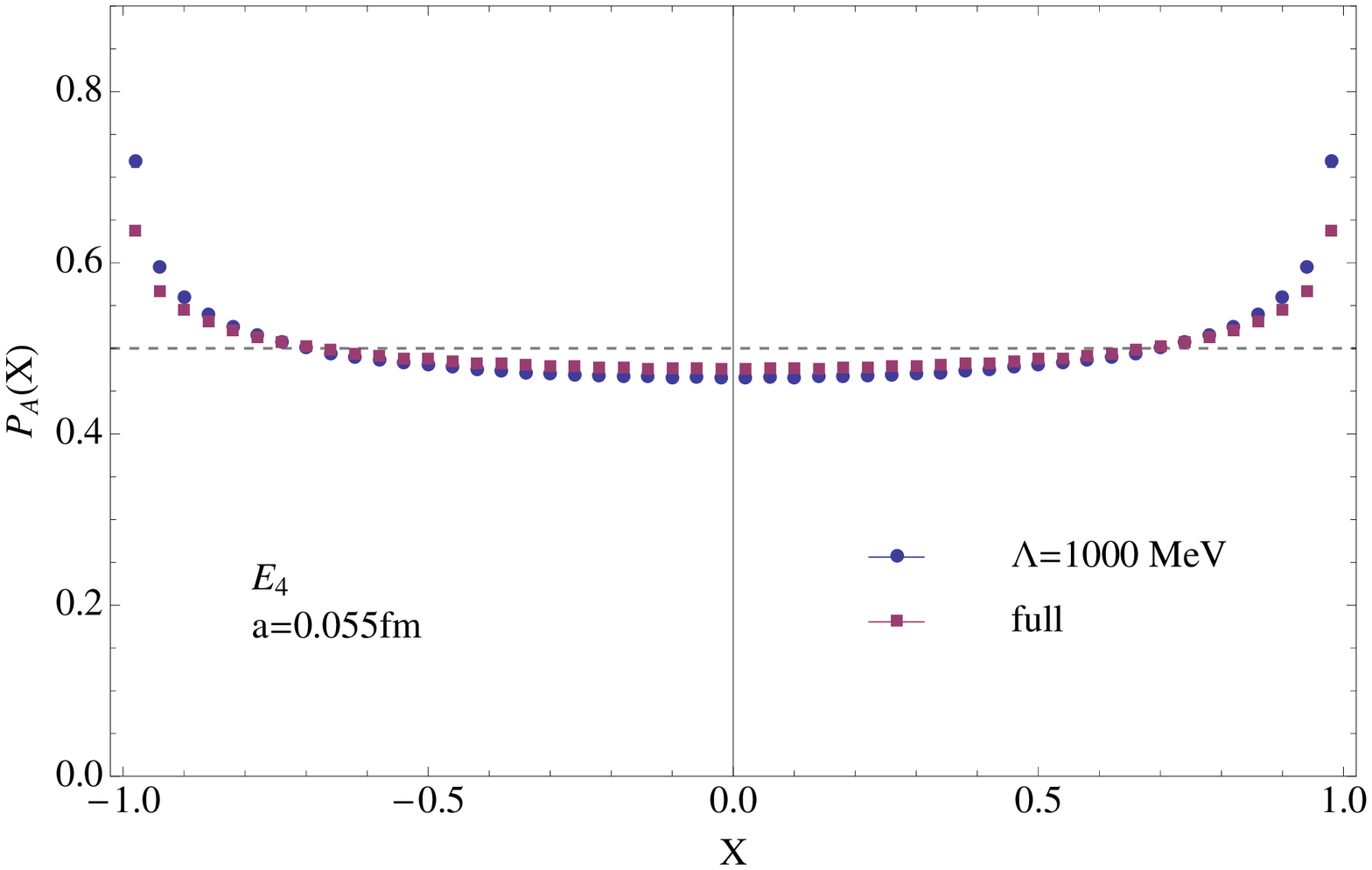}
     }
    \vskip 0.10in
    \centerline{
    \hskip 0.08in
    \includegraphics[width=8.8truecm,angle=0]{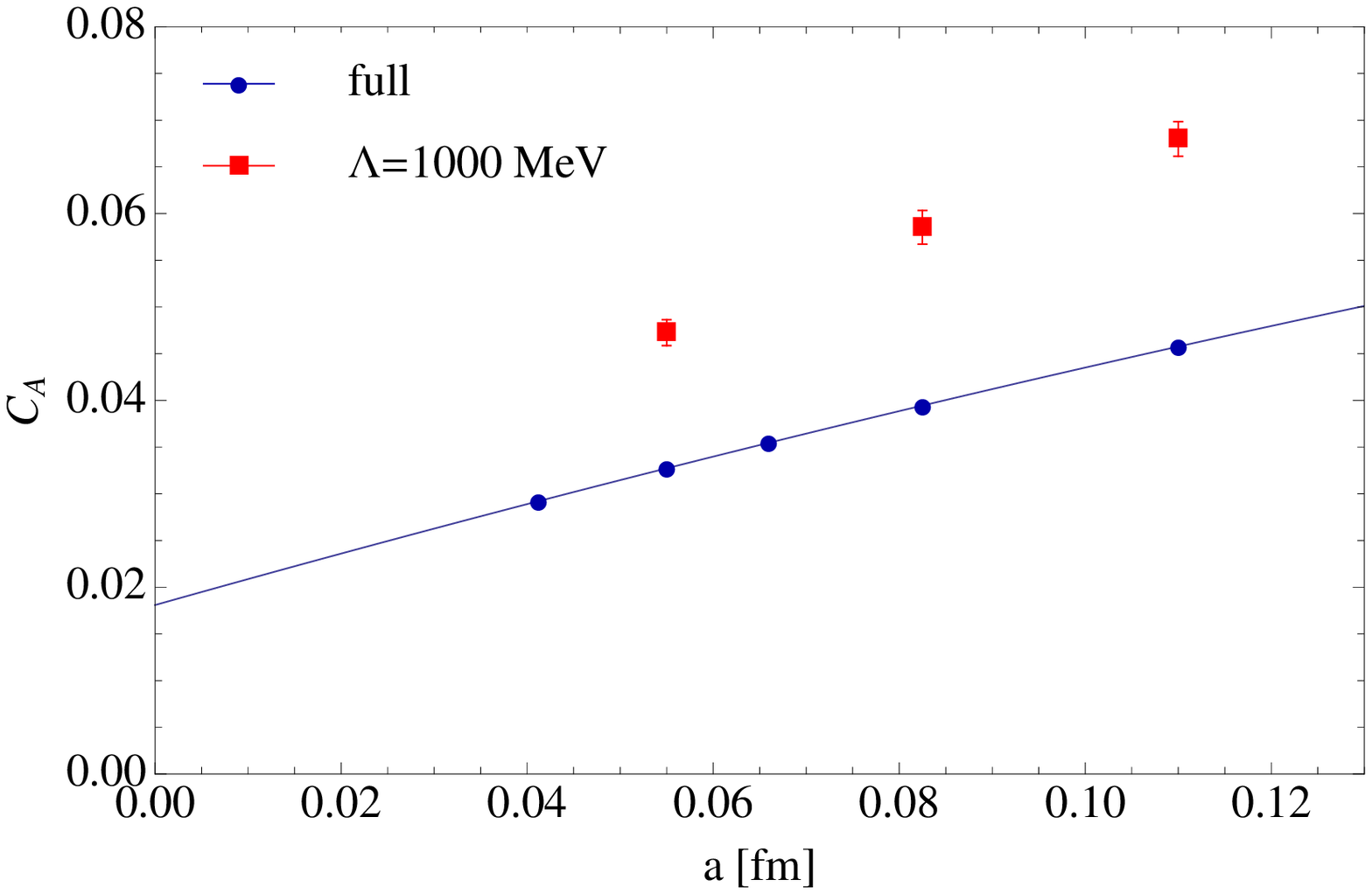}
     }
     \vskip 0.00in

     \caption{Comparison of dynamical polarization properties for field strength tensor
              $F$ and the effective field strength $F^{\Lambda}$, $\Lambda = 1000$ MeV.
              See discussion in the text.}
    \label{fig:PA_eff}
\end{center}
\end{figure}

\noindent{\bf 6. Effective Field Strength Tensor.}
It is revealing to compare dynamical polarization tendencies of fully fluctuating field
strength to those of effective fields defined by equation (\ref{eq:430}).
In Fig.\ref{fig:PA_eff} (top right) we show the comparison of absolute $\Xg$--distributions 
for $F$ and $F^{\Lambda}$, $\Lambda=1000$ MeV, in ensemble $E_4$. The construction of effective 
field strength in this case involved the inclusion of $47$ overlap near--zero modes on average. 
As one can see, while the dynamical polarization tendency has somewhat increased in the effective 
field, the two cases do not differ qualitatively at all. This happens despite the fact that
the associated reference $\Xg$--distributions, which are only kinematic, differ significantly 
(top left of Fig.\ref{fig:PA_eff}). In the bottom panel of Fig.\ref{fig:PA_eff} we added to 
the continuum extrapolation plot of Fig.\ref{fig:PA_all} the correlation 
coefficients $C_A$ at $\Lambda=1000$ MeV for ensembles with available eigenmodes. 
This shows that the correlation coefficients in the effective field are of comparable 
magnitudes to those for full field strength, and more so in the continuum limit.

Effective fields at fermionic scale $\Lambda$ suppress short distance fluctuations regardless 
of whether they contribute dynamically or represent pure noise. The above comparison thus speaks 
to the fact that absolute polarization measure has little sensitivity to the presence
of noise, as emphasized in the opening remarks, and to the fact that the dynamical
self--duality effect is mainly due to low--energy fields.

\bigskip
\noindent{\bf 7. Discussion.}
In this work we constructed dynamical measures of self--duality in the gauge field. 
These measures are based on the notion of absolute $\Xg$--distribution~\cite{Ale10A} 
which in turn represents a differential correlational characteristic of polarization. 
From the conceptual standpoint, this is an important step forward from initial approaches 
of Refs.~\cite{Hor01A, Gat02A} that are only kinematical, and thus arbitrary. Indeed, 
not even a sign of dynamical tendency can be deduced from such results.

Using the above framework, we performed a detailed study of self--duality in pure--glue SU(3) 
gauge theory using lattice regularization. Our results indicate that this theory involves 
positive dynamical tendency for self--duality, meaning that QCD dynamics produces enhancement 
of self--duality relative to statistical independence in duality components. However, the observed 
effect is very weak. Indeed, the measured correlation coefficient is only $\cop_A \approx 0.018$ 
while, for example, it is straightforward to prescribe dynamics for measured marginal distributions 
of duality components, that would enhance this correlation at least 20--30 times. It is significant 
in this regard that the same is true for low--energy effective fields obtained via eigenmode 
expansion of 
the field--strength tensor, despite of the associated ``noise reduction''. We are thus led 
to conclude that self--duality, and hence classicality, does not manifest itself as 
a significant feature of pure--glue QCD dynamics. This is consistent with previous 
arguments~\cite{WittenUA(1),Witt98A,Hor02B} as well as with lattice QCD 
studies~\cite{Hor02B,Hor03A,Hor05A, Ilg07A} indicating an intrinsically non--classical paradigm 
of QCD topological charge fluctuations.

\bigskip

\noindent{\bf Acknowledgments:} 
Andrei Alexandru is supported in part by U.S. Department of Energy under grant DE-FG02-95ER-40907.
Ivan Horv\'ath acknowledges warm hospitality of the BNL Theory Group during which part 
of this work has been completed.

\end{document}
\bye